\documentclass[reprint,
superscriptaddress,
amsmath, amssymb,
prd,
twocolumn,
nofootinbib]{revtex4-2}
\usepackage{color}

\usepackage{bm}
\usepackage[utf8]{inputenc}
\usepackage{makecell}
\usepackage{subfig}
\usepackage{graphicx}  
\usepackage{dcolumn}   
\usepackage{bm}        
\usepackage[english]{babel}
\usepackage{booktabs, multirow, array}
\usepackage[table]{xcolor}
\usepackage{float}
\usepackage{makecell}
\usepackage{mathtools}
\usepackage{amsmath}
\usepackage[colorlinks]{hyperref}
\definecolor{LinkColor}{rgb}{0.75, 0, 0}
\definecolor{CiteColor}{rgb}{0, 0.5, 0.5}
\definecolor{UrlColor}{rgb}{0.75, 0, 0.0}
\hypersetup{linkcolor=LinkColor}
\hypersetup{citecolor=CiteColor}
\hypersetup{urlcolor=UrlColor}

\usepackage[normalem]{ulem} 

\begin{document}

\title{\bf Bounds on tidal charges from gravitational wave ringdown observations}
\author{Akash K Mishra}
\email{akash.mishra@icts.res.in}
\affiliation{International Centre for Theoretical Sciences, Tata Institute of Fundamental Research, Bangalore 560089, India}
\author{Gregorio Carullo}
\email{gregorio.carullo@nbi.ku.dk}
\affiliation{Niels Bohr International Academy, Niels Bohr Institute, Blegdamsvej 17, 2100 Copenhagen, Denmark}
\author{Sumanta Chakraborty}
\email{tpsc@iacs.res.in}
\affiliation{School of Physical Sciences, Indian Association for the Cultivation of Science, Kolkata-700032, India}


\begin{abstract}
Black hole solutions in the braneworld scenario are predicted to possess a tidal charge parameter, leaving imprints in the quasinormal spectrum.
We conduct an extensive computation of such spectrum, and use it to construct a waveform model for the ringdown relaxation regime of binary black hole mergers observed by LIGO and Virgo.
Applying a Bayesian time-domain analysis formalism, we analyse a selected dataset from the GWTC-3 LIGO-Virgo-Kagra catalog of binary coalescences, bounding the value of the tidal charge.
With our analysis we obtain the first robust constraints on such charges, highlighting the importance of accounting for the previously ignored correlations with the other black hole intrinsic parameters.
\end{abstract}
\maketitle
\section{Introduction}\label{intro}

The fact that gravity is not a force, but rather the manifestation of curved spacetime has stumped generations of scientists.
The dynamics of gravity, namely the evolution of spacetime curvature, is determined by matter fields living in the spacetime, while the evolution of matter fields themselves follows from the spacetime geometry.
One way of implementing this concept is through the route taken by Einstein, equating the matter energy-momentum tensor with a geometrical quantity constructed out of the Ricci tensor and its trace, known as the Einstein tensor. 
The resulting equation is referred to as Einstein's equation, and the corresponding theory of gravity is named as General Relativity (GR).
However, many possible alternatives for achieving the same goal are conceivable.

Alternative theories of gravity, often referred to as modified gravity, achieve precisely this, by introducing modifications to Einstein's equations while attempting to address some of its limitations, which include the appearance of singularities in solutions of Einstein's equations, explaining late-time acceleration in cosmological contexts, among others.
The deviation from GR can be of various types~\cite{Berti:2015itd}, predominantly of the following nature --- (a) higher curvature corrections, e.g., Lovelock theories of gravity, (b) presence of additional spacelike dimensions, e.g., braneworld models and (c) non-minimal couplings between gravity and additional matter degrees of freedom, e.g., Horndeski theories. 
Regarding the matter content, several combinations of additional fields are conceivable, including exotic matter fields~\cite{Padmanabhan:2013xyr, DeFelice:2010aj, Sotiriou:2008rp, PhysRevLett.55.2656, Alvarez-Gaume:2015rwa, Maartens:2003tw, Kanti:2004nr, Dadhich:2000am, Chamblin:1999by, Chamblin:2000ra, Jacobson:2000xp, Kanti:1995vq, Sotiriou:2013qea, PhysRevD.68.104012, Alexander:2009tp}.

As evident from the above discussion, if not restricted by some fundamental principles, such possibilities of going beyond GR would be countably infinite. Thus, in what follows we will restrict ourselves to those theories, which adhere to these two principles --- (i) any modifications of GR must lead to a Lagrangian which maintains diffeomorphism invariance, i.e., the Lagrangian must be a scalar under arbitrary coordinate transformation, and (ii) the Lagrangians must give rise to second-order field equations, in order to avoid the Ostrogradsky instability~\cite{Woodard:2015zca}. In this work, we will discuss one such class of modified gravity theory, namely the braneworld scenario.

Besides theoretical restrictions, additional constraints on these modified theories of gravity can be established by testing their compatibility with observations from both the weak field and the strong field regimes of gravity. 
Constraints arise e.g. from Solar System and binary pulsar experiments~\cite{Will:2014kxa}, from the measurement of black hole (BH) shadow by the Event Horizon Telescope collaboration~\cite{EventHorizonTelescope:2019ggy}, and from measurements of gravitational waves (GWs) emitted from the merger of binary BHs and neutron stars~\cite{LIGOScientific:2016aoc, LIGOScientific:2017vwq, LIGOScientific:2021djp, Nitz:2021zwj}, through the LIGO-Virgo detectors~\cite{AdvLIGO, AdvVirgo}.
GWs originating from coalescing compact binaries, as detected by the LIGO-Virgo detectors, carry with them not only information about the source properties but also on the underlying theory of gravity. Consequently, the analysis of GW data serves also as an important tool to test various gravitational theories beyond GR~\cite{Berti:2015itd}.
Constraints on specific modified theories of gravity have already been derived using the analysis of inspiral data~\cite{Sennett:2019bpc,Nair:2019iur, Perkins:2021mhb} (for future perspectives see e.g. Refs.~\cite{Gupta:2020lxa, Perkins:2020tra, Toubiana:2020vtf, Klein:2022rbf}).
As evident, among all of these classes of observations, GWs probe the strongest and most dynamical gravity regime, thus presenting an exceptional opportunity to examine and test alternative theories in ways previously beyond our reach.

The ideal settings to test the gravitational interaction in the strong gravity regime are experienced around the most compact solutions of the gravitational field equations, namely BHs.
Intriguingly enough, despite being a solution of complicated field equations, BHs are the simplest objects in our universe, characterized by very few parameters.
In GR, this follows from the existence of no-hair theorems~\cite{Penrose1982SOMEUP,Ginzburg,Zeldovich,Zeldovich2,Israel,Carter,Hawking1972,Robinson,Bunting,Mazur,Mazur_review}, suggesting that BH families are fully parametrized by their mass, spin, and electric/magnetic charges, offering a remarkable simplification of our understanding of these cosmic objects.
In dynamical situations such as mergers, this generalises to the ``final state conjecture''~\cite{Penrose1982SOMEUP}, predicting that stationary BHs are a dynamical attractor for the remnant BH relaxing in realistic astrophysical scenario.
The no-hair theorems also get imprinted in the multipolar structure of the Kerr-Newman spacetime~\cite{Hansen:1974zz, Geroch:1970cd, PhysRevLett.26.331, Mazur:2000pn} through the fact that all the multipole moments of the Kerr-Newman geometry can be determined in terms of the mass, spin, and charges alone.
This elegant theorem implies that additional properties, e.g., an external scalar degree of freedom, do not leave any imprint on the BH's external characteristics. 
Although a BH can have electric/magnetic charge and there are several mechanisms that could endow BHs with such a U(1) charge (these include scenarios capable of explaining the dark matter content of our Universe), the lack of observations of magnetic monopoles, together with several known mechanisms through which charge can be quickly lost in astrophysical environments (see Refs.~\cite{Cardoso:2016olt, Carullo:2021oxn, Pereniguez:2023wxf, Dyson:2023ujk} and references therein) support the simplification that for astrophysical BHs electric/magnetic charge can be assumed to be negligible.
This assumption is also compatible with current GW observations~\cite{Carullo:2021oxn, Gupta:2021rod}.
Deviations from GR are however likely to introduce additional hairs to the BHs, which can then be detected using GW observations. 

Since these systems are dynamical, it is not straightforward to construct a test of the final state conjecture with an immediate interpretation. 
Perhaps the test with the most direct interpretation can be achieved by measuring  at late times (see~\cite{Baibhav:2023clw}) the BH quasi-normal mode (QNM) frequencies and damping times, dictating the relaxation phase of astrophysical BHs under external perturbations.
Intriguingly, both the oscillation frequencies as well as the damping time scales are unique functions of the BH ``hairs'', which for the Kerr BH are simply the mass and the spin.
The ringdown spectrum therefore serves as a distinctive signature of the underlying spacetime, distinguishing a Kerr BH from other solutions involving additional hairs.
As a result, analyzing the properties of the remnant object from the ringdown phase has emerged as a prominent research theme in the field of GW data analysis over the past few decades~\cite{Detweiler:1980gk, Echeverria:1989hg, Kokkotas:1999bd, Dreyer:2003bv, Buonanno:2006ui, Berti:2007fi, Berti:2009kk, Kamaretsos:2011um, Kamaretsos:2012bs, London:2014cma, Bhagwat:2016ntk, Baibhav:2017jhs, London:2018gaq, Baibhav:2018rfk, Brito:2018rfr, Carullo:2018sfu, Berti:2018vdi, Ota:2019bzl, Bhagwat:2019bwv, Bhagwat:2019dtm, Carullo:2019flw, Borhanian:2019kxt, Maselli:2019mjd, Cardoso:2019mqo, Cardoso:2019rvt, Giesler:2019uxc, Cabero:2019zyt,Hughes:2019zmt, Lim:2019xrb, Lim:2022veo, Bustillo:2020buq, JimenezForteza:2020cve, Mourier:2020mwa, Cook:2020otn, Laghi:2020rgl, Baibhav:2020tma, Okounkova:2020vwu, Carullo:2021dui, Capano:2021etf, Finch:2021iip, Isi:2021iql, Bhagwat:2021kwv, Ghosh:2021mrv, Carullo:2021oxn, Finch:2021qph, MaganaZertuche:2021syq, Ota:2021ypb, Forteza:2021wfq, Carullo:2021yxh, Li:2021wgz, Volkel:2022aca, Franchini:2022axs, Cotesta:2022pci, Isi:2022mhy,Forteza:2022tgq, Finch:2022ynt, Baibhav:2023clw, Ma:2023cwe, Franchini:2023eda, Bhagwat:2023jwv, Siegel:2023lxl, Pacilio:2023mvk, Nee:2023osy, Crisostomi:2023tle, Ma:2023vvr}.
In particular, through the measurement of the ringdown signals in binary BH merger events detected by the LIGO-VIRGO collaboration has also allowed to place observational constraints on multiple beyond-GR or beyond-Kerr scenarios~\cite{Laghi:2020rgl, Carullo:2021dui, Ghosh:2021mrv, Westerweck:2021nue, Carullo:2021oxn, Mishra:2021waw, Dey:2022pmv, Silva:2022srr}.
Interestingly, the ringdown signals of all the binary BH merger events are consistent with the Kerr-BH paradigm of GR.
One of the primary challenges in testing a modified gravity hypothesis with GW data lies in the lack of a full inspiral-merger-ringdown (IMR) waveform template for such theories, although significant progress has been achieved recently in this direction both on the analytical~\cite{Julie:2017pkb,Julie:2017ucp, Jain:2022nxs, Julie:2022qux, Jain:2023fvt,Yunes:2011we,Blazquez-Salcedo:2016enn,Blazquez-Salcedo:2017txk,Blazquez-Salcedo:2020caw,Cardoso:2019mqo,McManus:2019ulj,Pierini:2021jxd,Pierini:2022eim,Srivastava:2021imr,Wagle:2021tam,Cano:2019ore,Adair:2020vso,Cano:2020cao,Cano:2021myl,Cano:2023tmv,Cano:2023jbk,Cano:2023qqm,Li:2023ulk,Tattersall:2017erk,Franciolini:2018uyq} and numerical~\cite{East:2021bqk, Ripley:2022cdh, Corman:2022xqg, Evstafyeva:2022rve, Cayuso:2023aht} sides.
In this context, a relatively simpler approach involves focusing solely on the analysis of the ringdown segment of the signal.
On general grounds, this regime is expected to be morphologically similar to the GR case, with the black hole relaxation dominated by the fundamental quadrupolar mode (as explicitly verified within full numerical relativity in the similar Kerr-Newman case~\cite{Bozzola:2021elc}), hence amenable to accurate modeling within a perturbative framework. 
This approach provides a direct avenue for analyzing GW data with a waveform originating from modified gravity theories.

In this work, we consider a particular class of modification over GR due to the presence of an extra spatial dimension, not necessarily compact, known as the braneworld model~\cite{Maartens:2003tw, Perez-Lorenzana:2005fzz}. 
Within this framework, gravitational interaction can propagate not only within the familiar four spacetime dimensions, called the ``brane'', but also penetrate into the additional spatial dimension known as the ``bulk''.
In contrast to gravity, all the other standard model fields such as electromagnetism, and the strong and the weak nuclear forces, are confined on the brane, i.e. on the four-dimensional spacetime observers under considerations live in. 
The five-dimensional gravitational action is taken to be the Einstein-Hilbert action, while the gravitational field equations on the brane are obtained by appropriate projection of Einstein's equation on the bulk. 
This leads to additional correction terms to the four-dimensional Einstein's equations, proportional to the bulk Weyl tensor, while maintaining the second-order nature of the field equations. 
Notably, the effective gravitational field equations on the brane can be solved to yield a rotating BH solution on the brane, which resembles the Kerr-Newman solution, albeit with a significant difference. 
The rotating braneworld solution also involves a charge term, often referred to as the tidal charge, which originates from the bulk Weyl tensor and takes \emph{negative} values. 
The presence of such a nonzero and negative tidal charge in the BH solution can be regarded as a potential violation of GR and possible vindication of the existence of an extra spatial dimension.

Prior studies have already reported interesting findings regarding the constraints imposed on the tidal charge parameter from various observations, starting from the electromagnetic emission profile of accretion discs to the shadow of M87* and Sgr A* observed by the event horizon telescope~\cite{Chakravarti:2019aup, Banerjee:2019nnj, Horvath:2012ru, Zakharov:2018awx, Neves:2020doc, Banerjee:2019sae, Chakraborty:2021gdf}.  
In a previous work~\cite{Mishra:2021waw}, some of us have explored the consequences of this negative tidal charge term on the geometry of spacetime, specifically by examining its impact on the QNM spectra of BHs. 
However, the earlier analysis has a significant limitation --- it used the mass and spin measurements that were obtained assuming GR as the null hypothesis --- overlooking any potential correlations between the tidal charge and other remnant parameters, e.g., the mass and the spin of the BH. 
In this paper, we extend our previous analysis by employing a comprehensive Bayesian framework to examine the braneworld hypothesis against the GW ringdown data, including the entire correlation structure among all the free parameters. 
This means we consider all the hairs, namely the tidal charge, the mass, and the spin of the remnant BH as free parameters, and estimate them from the ringdown data directly, without assuming any connection to the GR values.   
We do so by numerically computing the gravitational QNMs of the braneworld rotating BH spacetime, and then modeling the ringdown waveform as the superposition of damped sinusoids and employing a full Bayesian analysis in the time domain.
With this setup, we will analyze ten events from the GWTC-3 catalog and obtain the first ever robust constraints on the tidal charge parameter derived directly from the GW data.

The rest of the paper is structured as follows: In Section~\ref{Section 2} we provide a concise overview of the rotating braneworld BH, discuss some of its interesting properties, and present the ringdown waveform constructed from a linear superposition of the QNMs. 
Section~\ref{Section 3} discusses our numerical framework, based on time-domain parameter estimation and the nested sampling algorithm.
In Section~\ref{Section 4}, we present the results of our parameter estimation analysis, including bounds on the tidal charge and Bayes Factors between the GR and the braneworld hypotheses. We conclude and discuss the implications of our findings in Section~\ref{Section 5}.

\emph{Notations and Conventions:} Throughout this paper we use the mostly positive signature convention, i.e. the four-dimensional flat spacetime metric in Cartesian coordinates can be expressed as, $\eta_{\mu \nu}=\textrm{diag.}(-1,+1,+1,+1)$. We use uppercase Roman indices $A,B,C,\ldots$ to describe the five dimensional spacetime coordinates, while Greek indices $\mu,\nu,\alpha,\ldots$ to describe four dimensional spacetime coordinates. We will work with $G=1=c$ units, but will restore these fundamental constants while comparing with the observed GW data. 
\section{Rotating Braneworld Black Hole and the Ringdown Waveform}\label{Section 2}

In this section, we provide a brief discussion of the rotating braneworld geometry and the field equation from which it originates, followed by the ringdown waveform construction using the QNMs. 

\subsection{Geometry on the brane}

We provide here a bird's eye view on the geometry of the braneworld scenario and the relevant equations which will be used in this paper. Further details have been extensively discussed in Ref.~\cite{Mishra:2021waw}, as well as in the excellent review \cite{Maartens:2003tw}. Thus for brevity, we will only provide a concise overview here. 

As emphasized earlier, the gravitational dynamics in the five-dimensional spacetime, known as the bulk, is governed by five-dimensional Einstein's equations,
\begin{align}
G_{AB}=8\pi G_{5}T_{AB}~,
\end{align}
where $G_{5}$ is the five-dimensional gravitational constant and $T_{AB}$ is the bulk energy-momentum tensor, which in general has the following form, 
\begin{align}
T_{AB}=-\Lambda g_{AB}+(T_{\mu \nu}-\lambda_{\rm b}g_{\mu \nu})e^{\mu}_{A}e^{\nu}_{B}\delta(y)~.
\end{align}
Here, $\Lambda$ is taken to be negative and corresponds to the bulk cosmological constant, $y$ is the extra spatial coordinate, and $e^{\mu}_{A}$ projects bulk quantities on the four-dimensional brane, located at $y=0$, with metric $g_{\mu \nu}$. 
Moreover, $T_{\mu \nu}$ is the matter energy-momentum tensor on the brane, and $\lambda_{\rm b}$ is the brane tension. 
Since observers lie on the brane hypersurface, the gravitational field equations on the brane is derived by projecting the five-dimensional Einstein tensor $G_{AB}$ onto the four-dimensional brane, which involves the utilization of the Gauss-Codazzi and Mainardi relations~\cite{Shiromizu:1999wj}.
These relations establish connection between curvatures of the bulk spacetime with that on the brane. 
This leads to the following effective gravitational field equations on the brane~\cite{Shiromizu:1999wj},
\begin{align}\label{brane_eqn}
~^{(4)}G_{\mu \nu}+E_{\mu \nu}=8\pi G_{4} T_{\mu \nu}+\frac{48\pi G_{4}}{\lambda_{\rm b}}\Pi_{\mu \nu}~.
\end{align}
Here, $^{(4)}G_{\mu \nu}$ is the Einstein tensor on the brane, $G_{4}$ is the four-dimensional gravitational constant and $E_{\mu \nu}=C_{ABCD}e^{A}_{\mu}n^{B}e^{C}_{\nu}n^{D}$ is the projected Weyl tensor on the brane. 
On the matter side, besides $T_{\mu \nu}$, there is an additional contribution $\Pi_{\mu \nu}$, which is quadratic in the matter energy-momentum tensor $T_{\mu \nu}$ on the brane. 

Since we are interested in vacuum solutions, there is no energy-momentum tensor on the brane, and hence the right hand side of the effective gravitational field equations on the brane goes to zero\footnote{Note that for neutron star the right hand side of Eq.~(\ref{brane_eqn}) remains non-zero and hence can be used to constrain the brane tension $\lambda_{\rm b}$ \cite{Chakravarti:2019aup}.}. 
Thus for vacuum brane, the effective gravitational field equations will be different from vacuum Einstein's equations by the term $E_{\mu \nu}$, constituting the correction over the Einstein tensor. 
Owing to the symmetry properties of the Weyl tensor, it follows that $E_{\mu \nu}$ is trace-less and hence it mimics the energy-momentum tensor of the Maxwell field, which is also trace-less in four spacetime dimensions.
Thus, one can replace $E_{\mu \nu}$ by the energy-momentum tensor of the Maxwell field and hence obtain the following rotating BH metric as the solution of Eq.~(\ref{brane_eqn}), with the following form,
\begin{align}
ds^2&=-\frac{\Delta}{\Sigma}(dt- a \sin^2\theta\, d\phi)^2+\,\Sigma\left[\frac{dr^2}{\Delta}+\,d\theta^2\right] 
\nonumber
\\
&\qquad+\frac{\sin^2\theta}{\Sigma}\left[a\,dt-(r^2+a^2)d\phi\right]^2~.
\end{align}
In the above metric, the quantities $\Delta$ and $\Sigma$ has the following definitions: $\Delta\equiv r^{2}+a^{2}-2Mr+q$ and $\Sigma\equiv r^{2}+a^{2}\cos^{2}\theta$. Here $(M, a, q)$ represents the mass, the spin and the tidal charge originating from the bulk spacetime, respectively. 
Despite the striking resemblance of the above spacetime metric with that of the Kerr-Newman solution in GR, it is important to note a crucial difference: the tidal charge parameter exclusively assumes negative values, in contrast to the U(1) Kerr-Newman charge, which always contributes positively to the metric components. 
This distinctive attribute is an intrinsic signature of the braneworld model, and forms the central aspect that we wish to study in this paper from a data analysis prospective. 
The above spacetime metric admits two horizons $r_{\pm}$, located at the zeroes of $\Delta$, yielding $r_{\pm}=M\pm\sqrt{M^{2}-a^{2}-q}$. 
Since the tidal charge parameter is negative, the extremal spin value is bounded by $(a/M)^{2}\leq 1-(q/M^{2})$, which can be larger than unity as well. 
This is another tantalizing feature of the braneworld scenario, namely the maximum attainable BH spin can exceed unity depending upon the value of $q$.
This will also play a key role in our subsequent analysis.

\subsection{Waveform model for the ringing black hole on the brane}

After the merger of two BHs on the brane, the newly-born remnant relaxes towards a stationary state, ``ringing-down'' according to the QNMs of the system, a process generated by the remnant distortions away from stationarity~\cite{Buonanno:2006ui}.
Far from the peak of the GW signal, the dominant GW mode is well described by the superposition of damped sinusoids with frequency and decay times corresponding to the QNMs associated with the linear perturbation of the system\footnote{Quadratic QNMs in comparable mass binaries were extracted from numerical simulations in Refs.~\cite{London:2014cma,Mitman:2022qdl,Cheung:2022rbm}. See references therein for the vast literature of second order BH perturbation theory. Recent developments not included in those references are Refs.~\cite{Bucciotti:2023ets, Perrone:2023jzq, Redondo-Yuste:2023seq}.}, which can be computed in the framework of BH perturbation theory.
The ringdown waveform, thus, can be modelled as~\cite{Berti:2005ys}, 
\begin{align}\label{waveform}
h_{+}(t)&+i \, h_{\times}(t)=\sum_{lmn}\Big[\mathcal{A}_{lmn} e^{i(\omega_{lmn} t + \phi_{lmn})} 
\nonumber
\\
&\qquad \times e^{-t/\tau_{lmn}}\, S_{lmn}(\iota, \phi) \Big]~.
\end{align}
Here $(\mathcal{A}_{lmn}, \phi_{lmn})$ represent the amplitude and the phase corresponding to each of the $(l,m,n)$ mode, where $n$ is the overtone number and $(l,m)$ are the angular indices. 
The overtone index arranges distinct modes in a descending sequence based on their decay times, such that $n=0$ mode is the longest living mode, with maximum decay time. 
$S_{lmn}$ represents the spin-weighted spheriodal harmonics, capturing the angular contribution to the waveform, with $(\iota, \phi)$ being the inclination and the azimuthal angle of the source with respect to the GW detectors, respectively. 
These were computed using Eqs.~(4-5) of Ref.~\cite{Kidder:2007rt}.
The time coordinate $t$ in the above expression starts from is $t=t_{0}$, where $t_0$ corresponds to the onset of the ringdown phase. 
While, the frequency and the decay time are given by $\omega_{nlm}=2\pi\, f_{lmn}$ and $\tau_{lmn}$, respectively. 
Note that, by expressing the waveform as above, we have absorbed the luminosity distance dependence, i.e. the $(M_f/D_l)$ pre-factor, into the amplitudes, where $M_{\rm f}$ is the mass of the merged BH and $D_{l}$ is the luminosity distance. 
The amplitudes of the $\pm m$ angular modes are related by $A_{l-mn} = A^*_{l-mn}$, while the frequency and the decay time are related by $f_{l-mn}=-f_{lmn}$ and $\tau_{l-mn}=\tau_{lmn}$.
Above, we have ignored contributions from the ``counter-rotating'' (retrograde) modes with $\textrm{Re}\,\omega_{\ell  -m n}< 0$, since they are strongly suppressed for large values of $(a/M)$ in non-precessing systems~\cite{Berti:2005ys,Lim:2019xrb,Li:2021wgz}.

\begin{figure*}[!tb]
\includegraphics[width=0.99\textwidth]{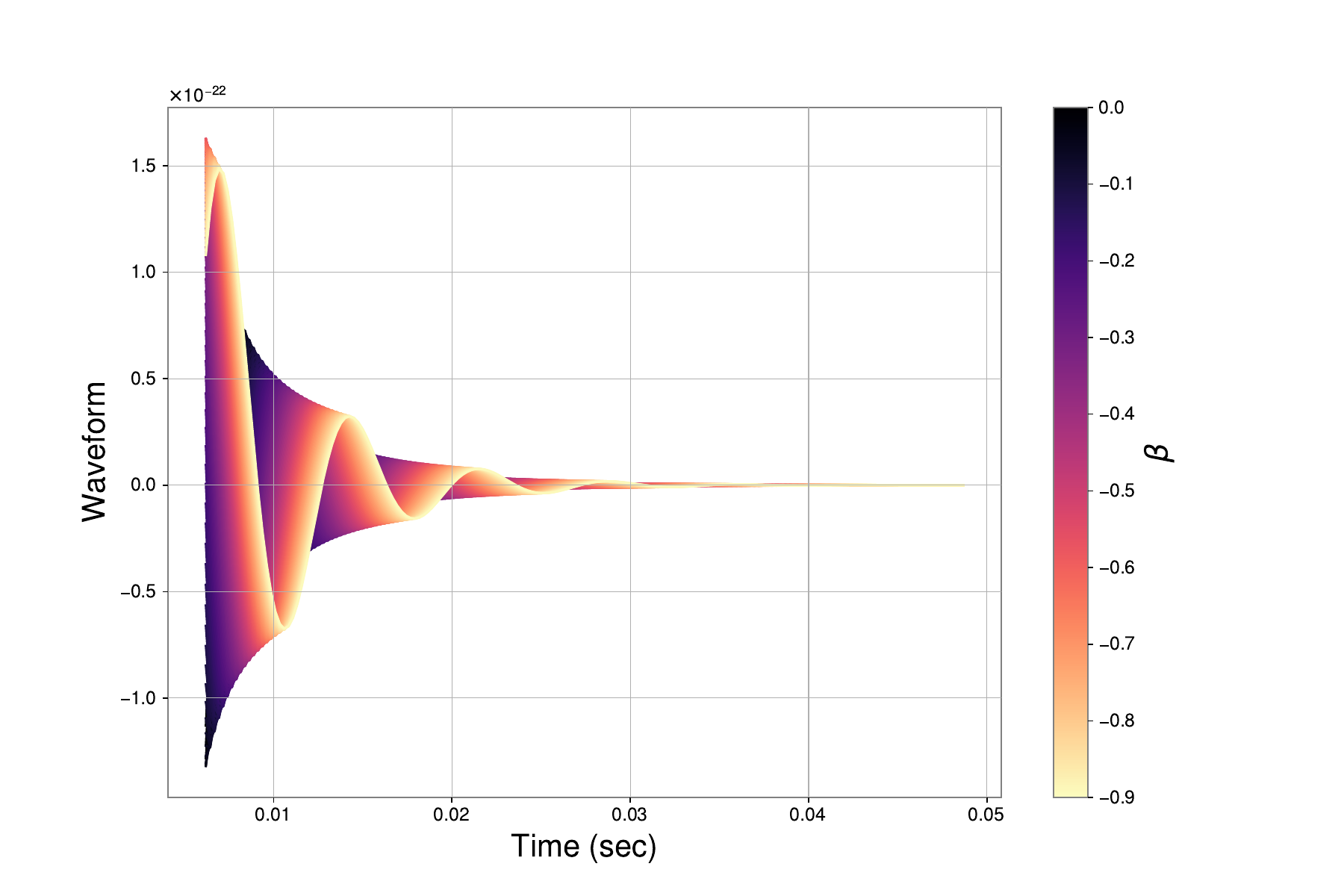}
\caption{Plus polarization of the ringdown waveform has been presented, which includes both the $(\ell,m,n)=(2,2,0)$, and $(2,2,1)$ modes, coloured as a function of the dimensionless tidal charge parameter $\beta\equiv(q/M^{2})$. 
As evident, when increasing the value of $|\beta|$, the oscillation frequency decreases, while the damping time increases. This result will form the main crux of our subsequent time domain analysis.}
\label{tidal_waveforms}
\end{figure*}

As discussed in Sec.~\ref{intro}, the frequency and decay time, and as a result the ringdown waveform, associated with the QNMs of the Kerr BH are uniquely determined by the mass and spin of the BH. 
However, in the presence of the tidal charge parameter, we have the frequency and the decay time scale to read $f_{lmn}=f_{lmn}(M_f,\chi,\beta)$ and $\tau_{lmn}=\tau_{lmn}(M_f,\chi,\beta)$.
These quantities determine how the ringdown waveform is impacted by the presence of the tidal charge. 

The determination of $(f_{nlm},\tau_{nlm})$ follows the following route --- (a) One first perturb the gravitational field equations on the brane and hence determine the equations satisfied by the radial and angular parts of the perturbed metric degrees of freedom. 
(b) Subsequently, one solves for both the radial and angular equations simultaneously and hence determine the QNM frequencies.
There are several numerical techniques available for solving the radial and angular perturbation equations, in this work we use the continued fraction method for solving them.  
One point needs to be elaborated here, which is regarding the perturbation of the effective gravitational field equations on the brane. 
The perturbation of the brane metric not only induces perturbations of the four-dimensional Einstein tensor, but also perturbation of the bulk Weyl tensor through $E_{\mu \nu}$. 
This poses a problem for separability of the perturbation equation into radial and angular parts. 
Though this is a serious hurdle for the U(1) charged Kerr-Newman solution, in the braneworld scenario it is not, since it follows that $\delta E_{\mu \nu}=(\ell/L)E_{\mu \nu}$, where $\ell$ is the length scale of the bulk geometry and $L$ is the length scale on the brane, satisfying $(\ell/L)\ll 1$. Thus, we can safely ignore any effect of higher dimensions on the perturbed field equations on the brane, except for contributions in the background geometry.
For an extensive discussion on this issue and also on the details of QNM computation in the braneworld scenario, we refer the reader to \cite{Mishra:2021waw}. 

For convenience, in this work we will work in units of $M$, i.e., we will set $M = 1$. Therefore, the re-scaled spin and the tidal charge can be defined as $\chi\equiv(a/M)$ and $\beta\equiv(q/M^2)$, respectively. 
To demonstrate the impact of the tidal charge on the GW waveform, in Fig.~\ref{tidal_waveforms} we depicted how the ringdown waveform originating from a braneworld BH changes, as the tidal charge parameter $\beta$ is varied. First of all, the variation of the GW waveform with respect to $\beta$ is precisely the opposite compared to the U(1) charge \cite{Carullo:2021oxn}. Moreover, as $\beta$ becomes more and more negative, i.e., the size of the extra spatial dimension increases, the GW ringdown waveform becomes more advanced in time. Also, with an increase of $|\beta|$, the oscillation frequency $f_{lmn}$ decreases, while the damping time $\tau_{lmn}$ increases. Thus there is a distinct signature of the tidal charge parameter on the phase of the GW ringdown waveform and this forms the main basis of our subsequent time domain analysis of the waveform to constrain this extra dimensional `charge'.  

The strain time series measured at the GW detector is a linear combination of both the plus $(h_{+})$ and cross $(h_{\times})$ polarizations, along with detector response functions, such that, 
\begin{equation}\label{projection}
h(t)=F_{+}\,h_{+}(t)+F_{\times}\,h_{\times}(t)~,
\end{equation}
where $F_{+}=F_{+}(\alpha, \delta, \psi)$ and $F_{\times} = F_{\times}(\alpha, \delta, \psi)$ are the detector response functions associated with the two polarizations. Both of these response functions depend on the parameters $(\alpha,\delta,\psi)$, which are the sky localization parameters, right ascension, declination and polarization angle, respectively. 
Furthermore, we restrict the ringdown template to a superposition of two quadrupolar modes, the fundamental ($l=2, m=2, n=0$) mode and its first overtone ($l=2, m=2, n=1$).
Although contributions from the overtones close to the waveform peak~\cite{Giesler:2019uxc, Isi:2019aib} have been shown to not correspond to any physical vibrational frequencies of the underlying spacetime~\cite{Baibhav:2023clw}, their usage allows to fit the waveform close to the peak with sufficient accuracy for current (low-SNR) GW signals.
Hence, we exploit them as effective terms that allow to push the analysis to earlier times, capturing more power present in the signal.
Their effectiveness in this sense is due to: the small lifetime of these contributions, which can then easily match short-lived nonlinear features near to the waveform peak; the lower frequency compared to the fundamental mode one, which more easily matches the merger frequency, also smaller than the fundamental mode.
The model parameters ($\theta$) that we wish to recover from data includes intrinsic parameters such as mass, spin, tidal charge, amplitudes, phases and also extrinsic parameters, i.e., $\theta = \{M_f, \chi, \beta, A_{lmn}, \phi_{lmn}, \alpha, \delta, \psi, \iota, \phi\}$. 
Note that in our notation $M_f$ represents the detector frame final mass, which is related to the source frame mass by a factor of $(1+z)$, where $z$ is the redshift of the GW source. 
Having discussed all the basic aspects of the background spacetime, as well as the waveform modeling, we will now proceed for Bayesian data analysis in the time domain. 

\section{Time-domain Bayesian data analysis}\label{Section 3}

Having briefly described the braneworld model and the  in the preceding section, we now move to the description of our data analysis framework. 
Bayesian analysis is a robust methodology, serving both as an effective tool for extracting information from data and also as a means to rigorously evaluate and compare different models. 
Conventional data analysis within the LIGO framework are performed in frequency domain, employing the complete inspiral-merger-ringdown waveform models. 
However, for situations in which we are interested only in a portion of the full signal, standard frequency domain approaches are not applicable, as they would lead to spectral leakage when Fourier transforming to frequency domain induced by the abrupt cut in the data. 
Therefore one must either construct a more involved likelihood function canceling the unwanted contributions~\cite{Capano:2021etf, Zackay:2019btq} or resort to a time-domain framework, where no need to apply a Fourier transform arises, at the price of having to deal with a non-diagonal likelihood. 
We employ here the latter approach.
In this section we start by discussing the fundamental components for the likelihood construction in the time-domain. 

\subsection{Time-domain likelihood}

We consider detector data composed of the ringdown signal $h(t)$ and an accompanying noise component $n(t)$, so that the time-domain detector data correspond to $d(t)=h(t)+n(t)$~\cite{DelPozzo:2016kmd,Carullo:2019flw, Isi:2019aib}.
For a thorough introduction to the topic, see Ref.~\cite{Isi:2021iql} and Chapter 3 of Ref.~\cite{PhD_thesis_Gregorio}.
We model the noise in the detector as a wide-sense stationary Gaussian process, hence fully characterized by its covariance matrix $\mathcal{C}_{ij} = \left<n_i n_j\right>$, since the mean can be easily set to zero without any loss of generality\footnote{Here, $n_{i}$ implies the noise at time $t=t_{i}$. This is because the numerical computation is performed over a sufficiently dense and discretized time grid.}.
For such stochastic processes, as the GW detection, the noise covariance matrix takes a symmetric Toeplitz form, given by
\begin{equation}
C_{ij} = \rho(|i-j|) , \quad\quad  \rho(k) \propto \sum_i n_{i}\, n_{i + k}~,
\end{equation}
where $\rho(k)$ is the auto-correlation function (ACF). 
In general, the ACF can be computed by correlating a long stretch of detector noise or from the inverse Fourier transform of the power spectral density (PSD).
Using the covariance matrix defined above, the time-domain log-likelihood can be expressed as, 
\begin{equation}
\ln\, \mathcal{L}(d|\theta) = -\frac{1}{2}\sum_{i,j = 0}^{N-1}\left[d_i - h_i(\theta)\right] C_{i j}^{-1} \left[d_j - h_j(\theta)\right]~.
\end{equation}
Here, $N$ corresponds to the number of grid points along the time axis and $\theta$ collectively denotes all the intrinsic and extrinsic parameters associated with the system. 
The likelihood, as defined above, forms the core of our analysis, allowing to isolate the post-merger signal and offering a convenient approach for analyzing the ringdown-only data. 
For a network of detectors, the joint likelihood is given by the product of individual likelihoods. 
The Bayesian posterior probability distribution $p(\theta|d)$ on the model parameters $\theta$ is related to the likelihood $\mathcal{L}(d|\theta)$, defined above, such that, 
\begin{equation}
p(\theta|d)=\frac{\mathcal{L}(d|\theta)\;\pi(\theta)}{\mathcal{Z}}\,,
\quad \mathcal{Z}=\int d\theta \mathcal{L}(d|\theta)\,\pi(\theta)~,
\end{equation}
where, $\pi(\theta)$ is the prior and $\mathcal{Z}$ represents the Bayesian evidence, quantifying the relative support that the observed data provide for different competing hypotheses (in our case, GR vs the braneworld scenario).
The Bayes factor, derived from the ratio of two Bayesian evidences, directly quantifies the strength of evidence in favor of one model over another, given by
\begin{equation}
\mathcal{B}^{\rm Kerr}_{\rm brane}=\frac{\mathcal{Z}_{\rm Kerr}}{ \mathcal{Z}_{\rm brane}}~.
\end{equation}
We will quote the above Bayes factor to provide a comparison between the Kerr BH in GR and the braneworld BH using real data from ten binary BH merger events.  

\subsection{Details of analysis and parameter estimation}

Having described the essential ingredients of our methodology, we now analyse the LIGO-Virgo data provided by the GW Open Science Center\cite{KAGRA:2023pio, Vallisneri:2014vxa}. 
This is performed completely within the context of the time-domain Bayesian framework described above, enabling us to rigorously assess the braneworld hypothesis against GR through the available GW datasets. 
{We start by band-passing $4096 s$ of data sampled at 4096 Hz around the merger time with a fourth order butterworth filter between frequency range $[20, 2047]$.
The ACF computation is obtained through the `FFT' method implemented in the \texttt{pyRing} software~\cite{pyRing}.
We work with $0.2$ seconds of ringdown data following the start time $t_0$. 
Given the ACF, we can compute the likelihood, and the last remaining ingredient is to employ an efficient method to extract samples from the posterior distribution. 

We employ the \texttt{DynamicNestedSampler}, which is a stochastic sampler of \texttt{dynesty}, a nested sampling algorithm library that excels in handling complex parameter spaces and provides efficient exploration of high-dimensional likelihood landscapes. 
We utilized the `unif' sampling method of \texttt{dynesty} to explore the parameter space and executed the sampling with a set of 3000 live points.
Furthermore, it is important to highlight that, in analyses involving ringdown-only data, the precise value of the ringdown start time ($t_0$, defined to be the maximum of the waveform peak~\cite{LIGOScientific:2020tif}) can in principle have a non-negligible impact on the analysis, and one should marginalise over the uncertainty on this parameter.
This is especially true when targeting short-lived features, such as overtones detection~\cite{Cotesta:2022pci}.
However, in our analysis the mode content is fixed, and we are only interested in comparing a signal with/without tidal charge contribution.
In this case, the impact of the start time uncertainty is much smaller, and to reduce computational cost we can fix it to a single value.
We repeat the analysis at two different times, $t_0 = [0, 10] \, M_f$ after the waveform peak.
The first time corresponds to a regime where non-linearity in the ringdown signal still play a role, and additional terms such as effective overtone components are required to correctly fit the signal~\cite{Baibhav:2023clw}.
Instead, for the second value of $t_{0}$, at current sensitivity, the signal can be well-described by contributions corresponding to the linear QNMs.
We project the complex-valued strain in Eq.~(\ref{waveform}) onto each LIGO detector by means of the corresponding antenna pattern functions, Eq.~(\ref{projection}), computed using the \texttt{LALSuite} library~\cite{lalsuite}.
In our analysis, we work with fixed values of right ascension $\alpha$, and declination $\delta$, employing maximum likelihood values derived from a comprehensive IMR analysis. 
This is done to avoid the template from latching to pre-merger portions of data and was verified to not introduce biases at the current detector sensitivity~\cite{LIGOScientific:2021sio}.
We also set the azimuthal angle $\phi = 0$, since it is completely correlated with the mode phases. 
We use uniform prior distributions on all the parameters, with ranges wide enough to encompass the full width of the posterior. 
Specifically, the prior distribution on the tidal charge parameter is uniform in the interval $(-1, 0)$. 
Further, we provide a conditional uniform prior on the spin parameter such that the extremality condition $(1-\chi^2-\beta>0)$ is satisfied for each sample point.
As a further check, we compared our results with the ones obtained using \texttt{pyRing}~\cite{pyRing} the standard analysis pipeline for ringdown-only analyses employed by the LVK collaboration, finding agreement.

\begin{figure*}[!tb]
\centering
\includegraphics[width=0.48\textwidth]{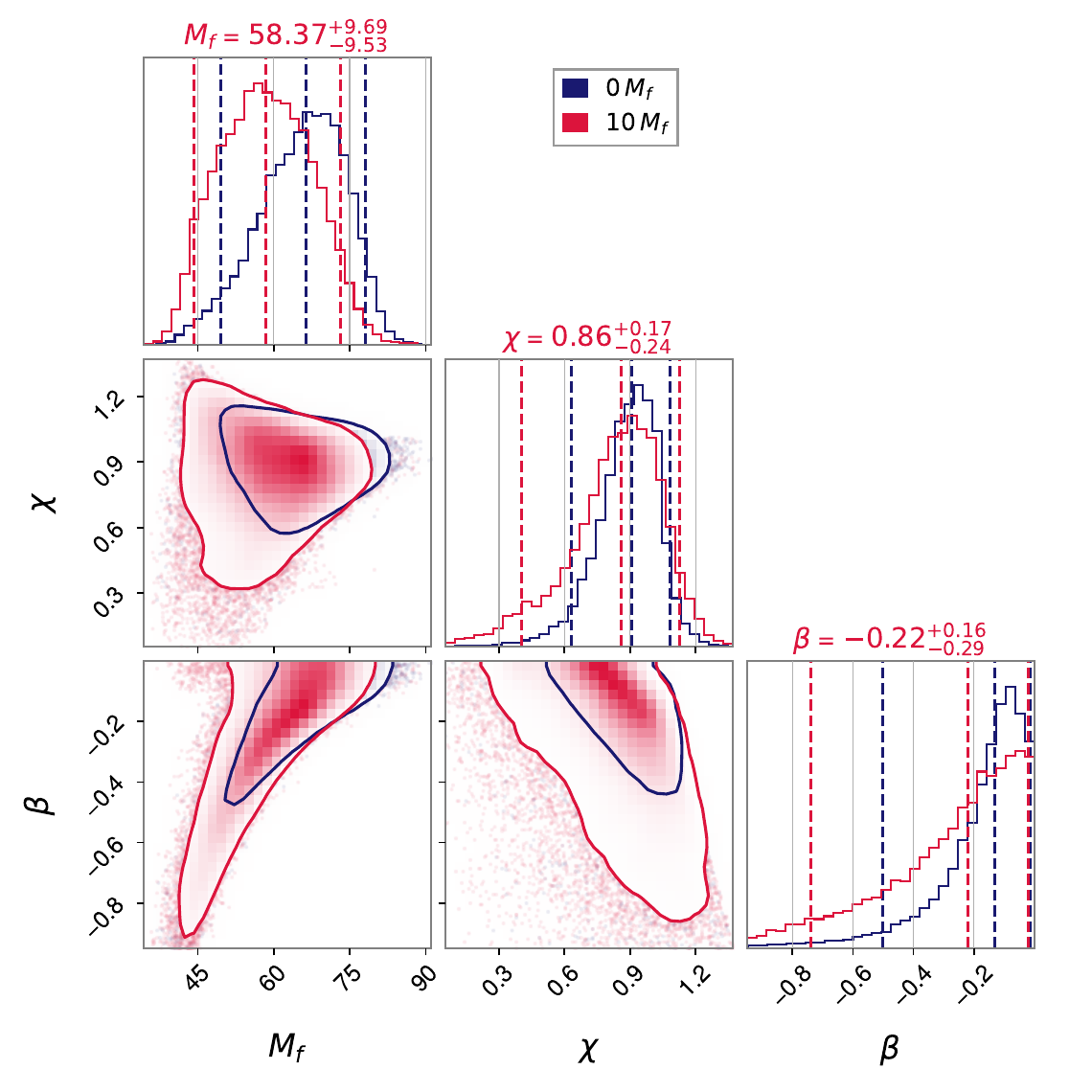}
\includegraphics[width=0.48\textwidth]{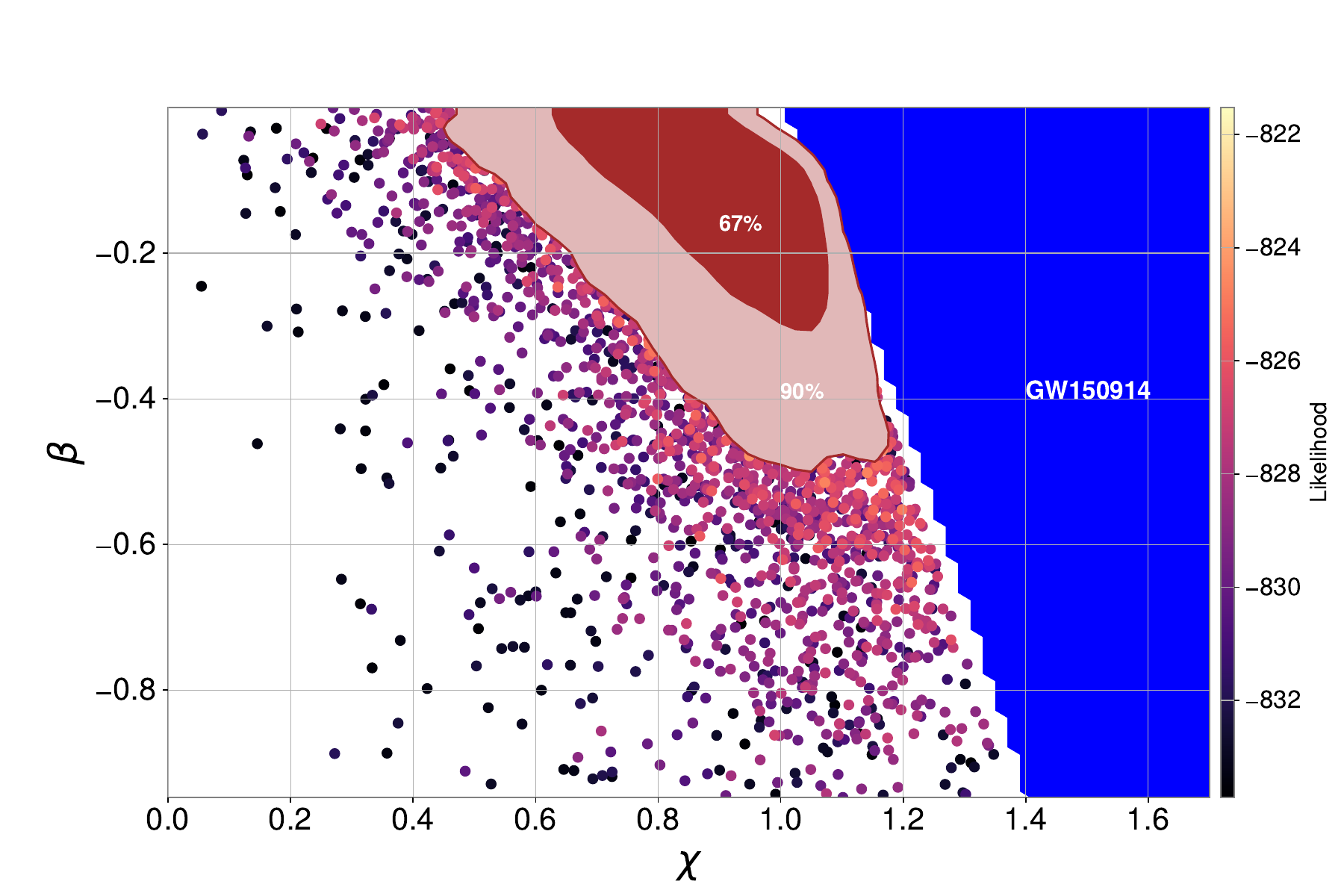}
\caption{
Left panel: posterior distribution for the remnant mass, spin, and tidal charge parameter resulting from the analysis of GW150914's ringdown signal. The contour lines delineate the $90\%$ credible regions.
Right panel: 2D posterior distribution of the tidal charge parameter and the final spin, for $t_0 = 0\, M_{f}$. The shaded blue area highlights regions in the parameter space where charge-spin values surpass the extremal limit.}
\label{GW150914_result}
\end{figure*}

\section{Constraints on the tidal charge}\label{Section 4}

The tidal charge parameter is intricately linked to the characteristics of the bulk Weyl tensor and presents a valuable avenue for discerning the strength and presence of additional spatial dimensions. 
The primary objective of this section is to conduct a comprehensive comparative analysis between the braneworld hypothesis and empirical ringdown data. 
Through this rigorous examination, we aim to obtain possible bound on the tidal charge parameter. 
It is important to note that, for a dedicated analysis focused solely on the ringdown phase, it is imperative that the selected event exhibits a substantial Signal-to-Noise Ratio (SNR) within this specific phase. 
This criterion ensures that our investigation can effectively capture the details of the model under investigation.
We start with an in-depth analysis of GW150914, the first gravitational-wave event detected, and subsequently extend our analysis to events in the third Gravitational Wave Transient Catalog (GWTC-3).
For computational reason, we restrict our analysis to a subset of events with a loud ringdown signal~\cite{LIGOScientific:2020tif} and substantial remnant masses, whose list can be found in Table~\ref{gwtc_2_table}.

\textit{Analysis with GW150914 ---} GW150914~\cite{LIGOScientific:2016aoc} originated from the merger of two BHs, with component masses approximately 36 and 29 times that of the sun, located at a distance of about $450$ Mpcs from Earth. 
Aside from being the first event to be detected, it is still among loudest within the GWTC-3 dataset, hence being particularly suitable for a ringdown-specific analysis. 
Let us start by investigating the braneworld hypothesis with this event. 
As mentioned earlier, to study both regimes where the signal still has non-negligible contributions from nonlinearities and the one where the dynamics is sufficiently close to linear perturbations, we consider two scenarios: one where $t_0$ is set to 0 (i.e. the peak of the waveform), and another where it is positioned at $10, M_f$ after the peak. 
In both cases, we model the GW signal using a linear combination of the $(2,2,0)$ and $(2,2,1)$ modes (the latter allowing to effectively describe the nonlinearities of the signal up to the peak at current SNR). 
We apply uniform priors on all the parameters as: $M_f \in [20, 150] M_\odot$, $\beta \in [0.0, -1.0]$, $\chi \in [0.0, \sqrt{1 - \beta}]$, $A_{220}, A_{221} \in [10^{-24}, 10^{-19}]$, $\phi_{220}, \phi_{221} \in [0, 2\pi]$, and $(\iota, \psi) \in [0, \pi]$. 
Note that the prior distribution on the remnant spin is dynamic in nature. 
In other words, the sampling process is constrained to select points up to the extremal value, as determined by the corresponding tidal charge sample point.
Employing the above described framework, and assuming the remnant object to be a braneworld BH, we conduct a comprehensive analysis of the ringdown data from GW150914. 
The findings are visually represented in Fig.~\ref{GW150914_result}. 
The left panel presents a three-dimensional corner plot illustrating the relationships among $(M_f, \chi, \beta)$ posterior samples for the different start times.
The contours delineate the $90 \%$ credible region. 
In the right panel we have shown the two-dimensional posterior distribution for ($\chi, \beta$), including the excluded region due to the extremality condition. 
\begin{figure*}[!tb]
\centering
\includegraphics[width=0.49\textwidth]{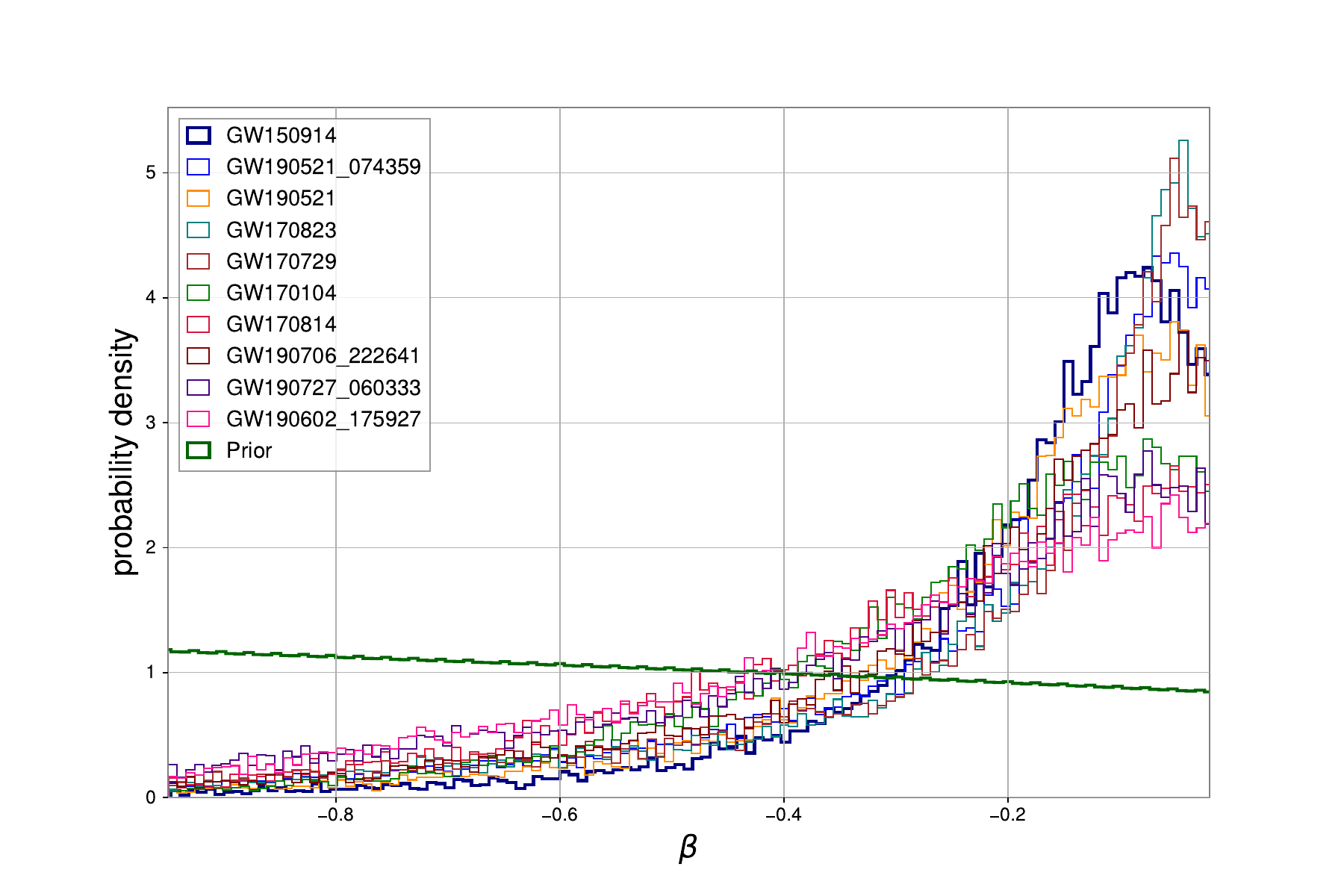}
\includegraphics[width=0.49\textwidth]{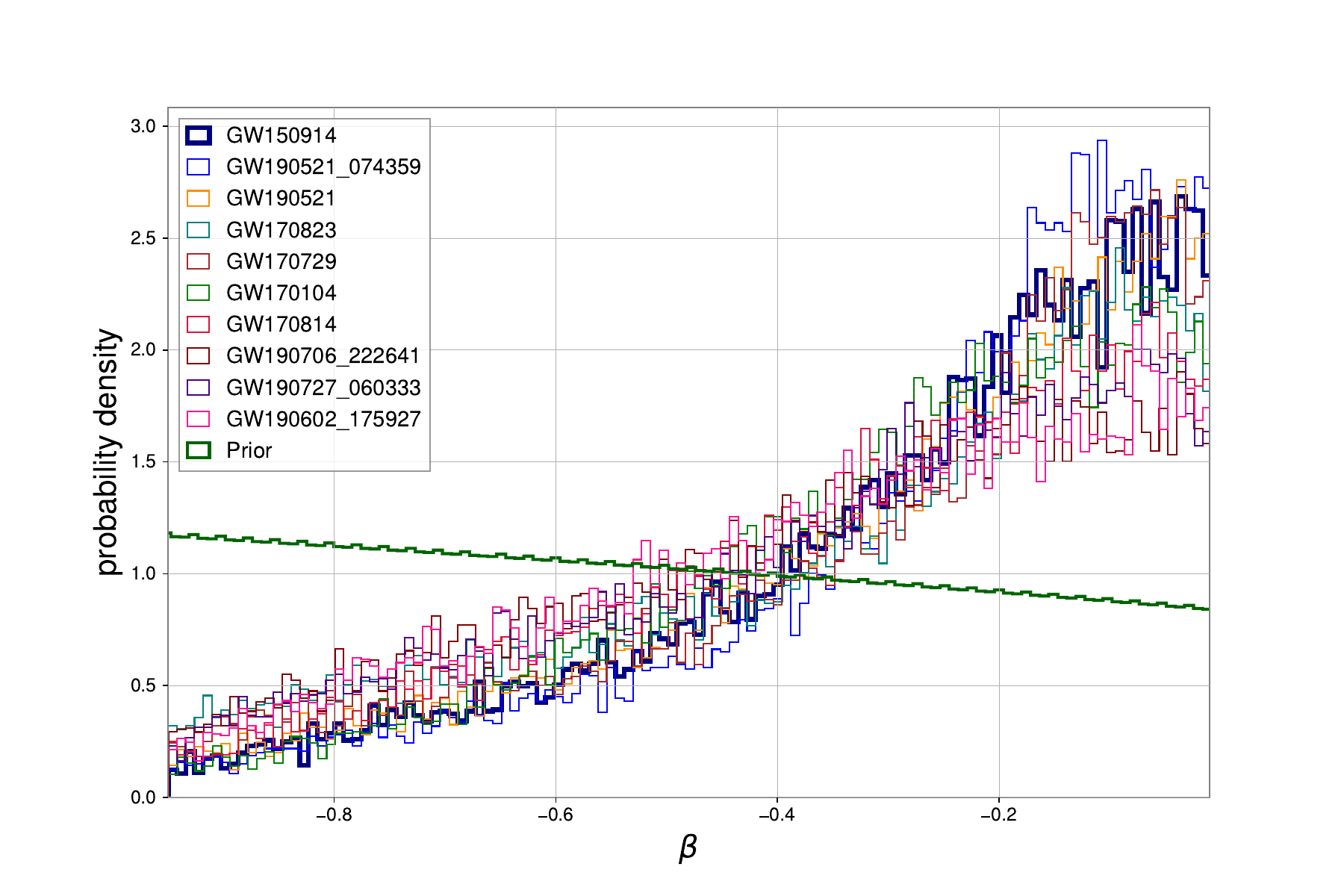}
\caption{Posterior distribution of the dimensionless tidal charge parameter ($\beta$) for 10 events from GWTC-3. The left and right panel corresponds to analyses with different ringdown start times, i.e.,  $t_0 = 0\, M_f$ and $t_0 = 10\, M_f$ after the peak. 
The green line represents the prior, when constrained to respect the extremality bound.}
    \label{GWTC_2_results}
\end{figure*}
\begin{figure*}[!tb]
\centering
\includegraphics[width=0.48\textwidth]{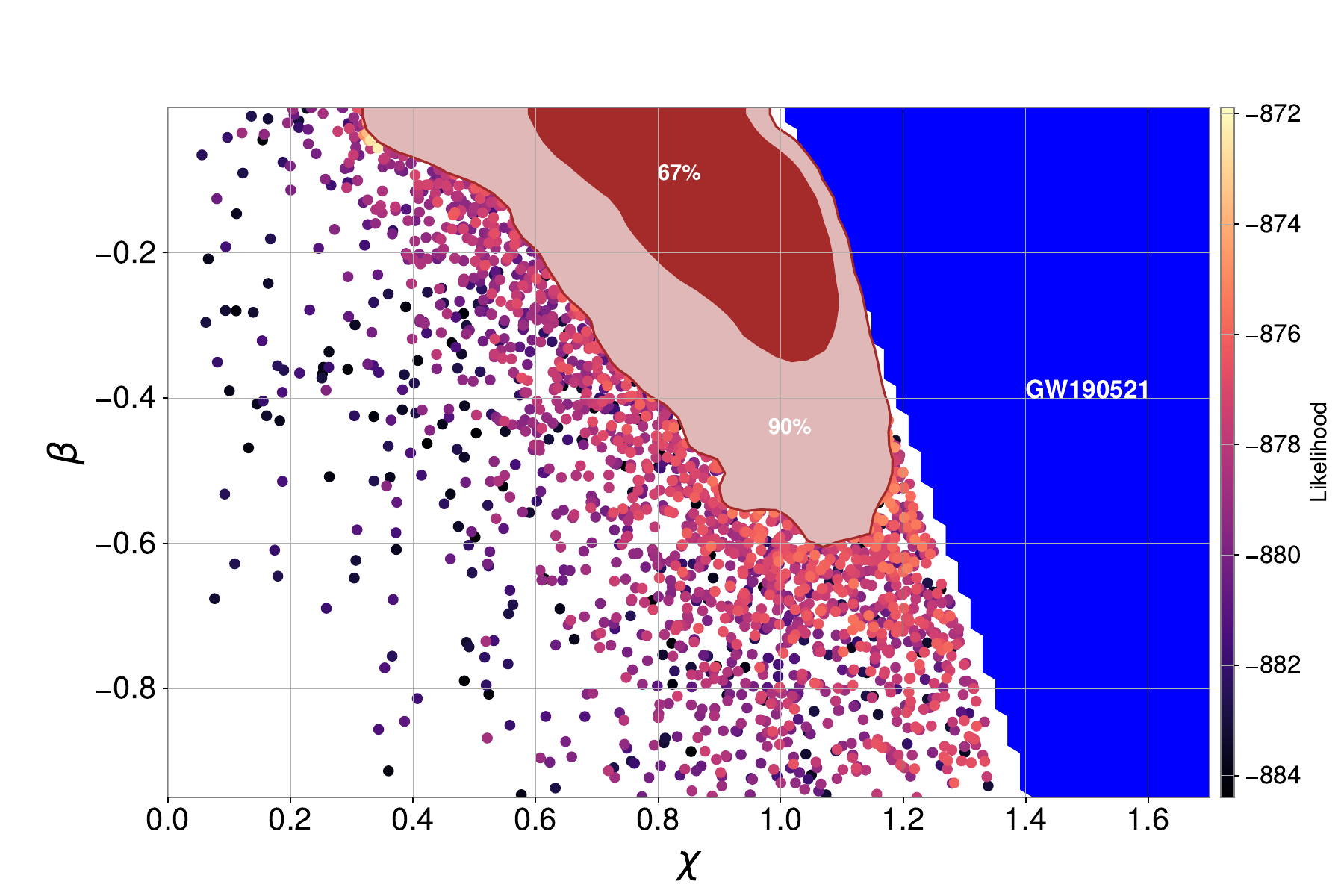}
\includegraphics[width=0.48\textwidth]{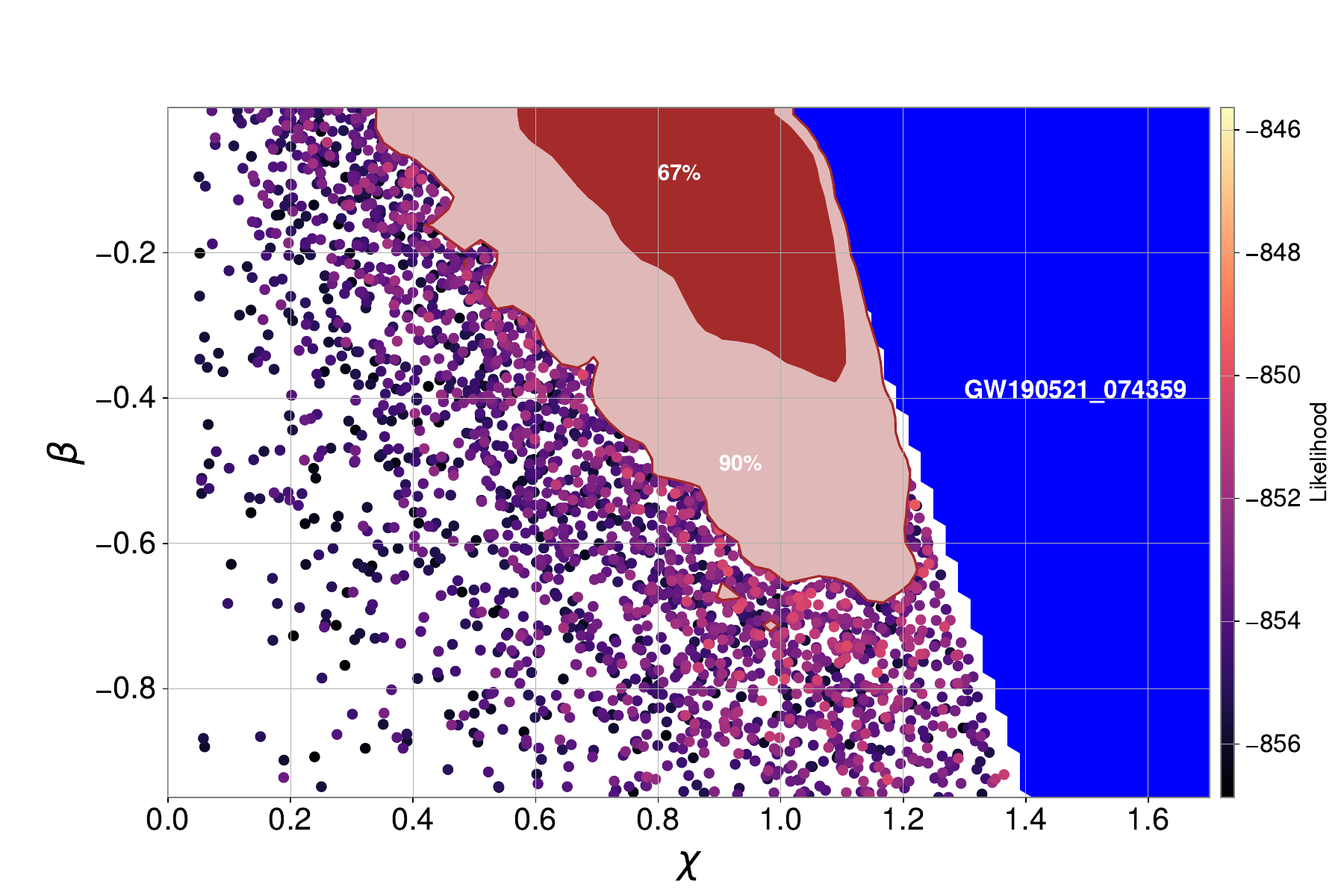}
\\
\includegraphics[width=0.48\textwidth]{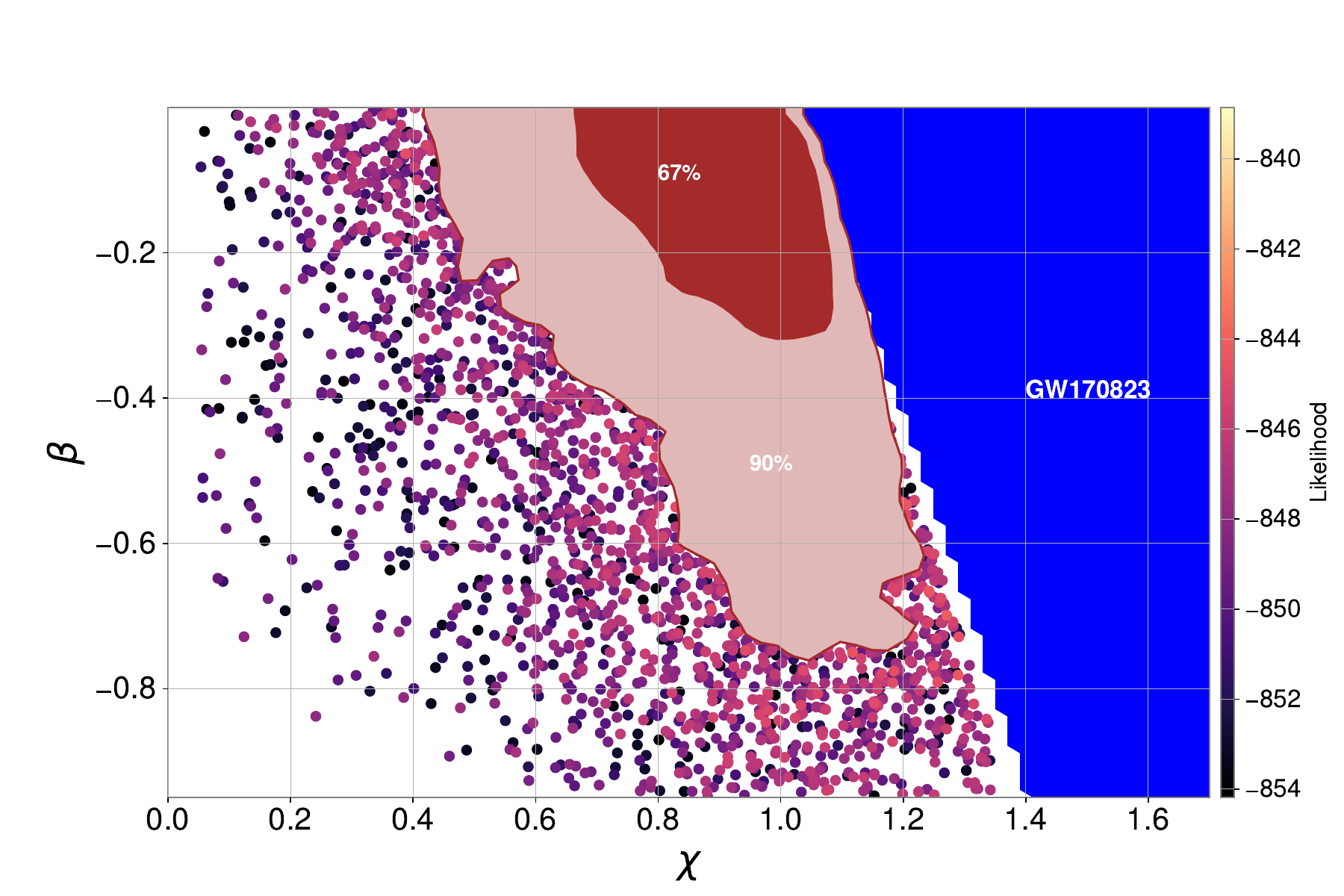}
\includegraphics[width=0.48\textwidth]{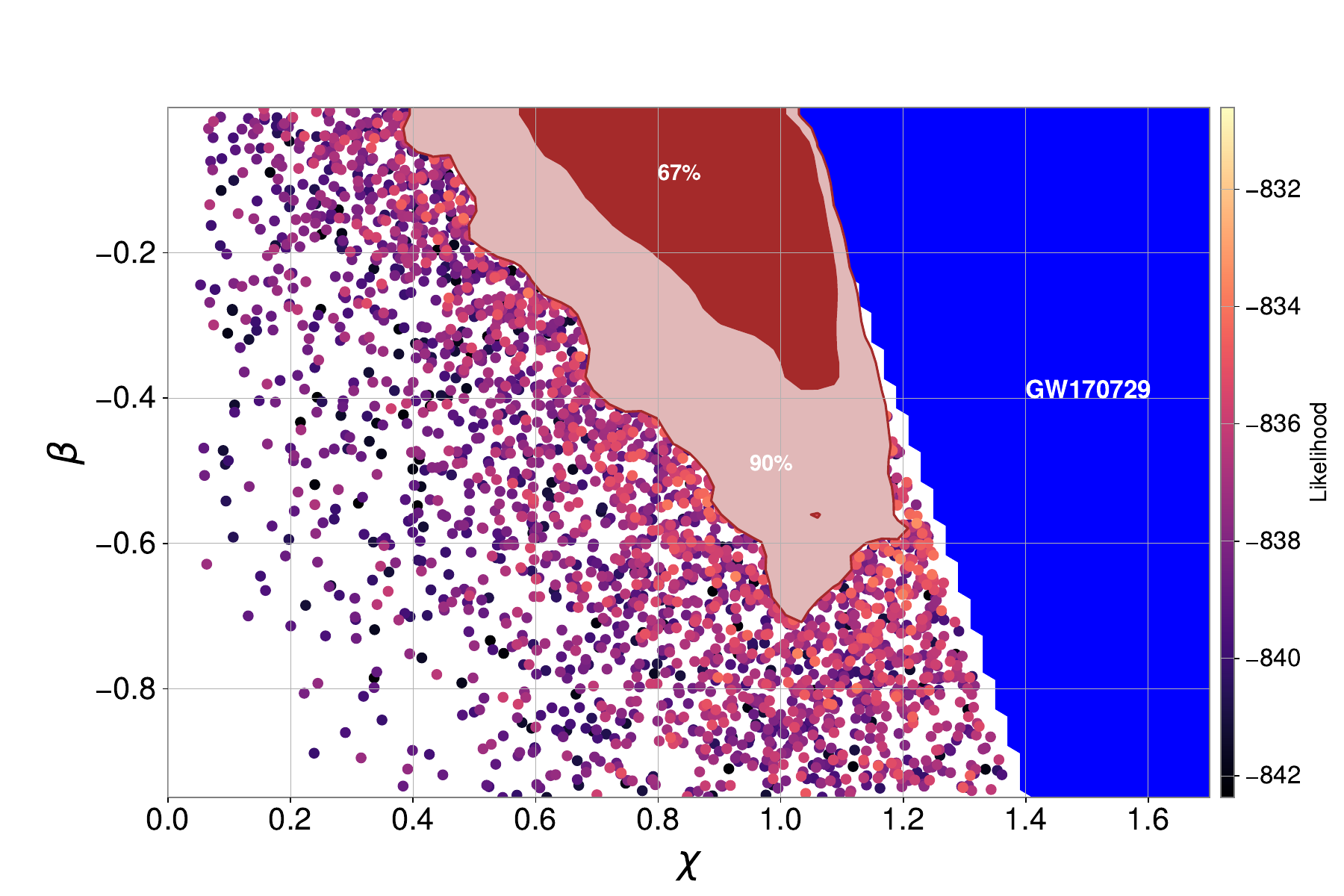}
\caption{Two-dimensional posterior distribution of final spin and dimensionless tidal charge parameter for different events, with $t_0 = 0 M_f$, where contours correspond to the $67\%$ and $90\%$ credible regions. The shaded blue area highlights regions in the parameter space where charge-spin values surpass the extremal limit.}
    \label{q_af_2d_posterior}
\end{figure*}
For these choices of ringdown start time, the resulting estimates for the tidal charge parameter are found to be $\beta = -0.13^{+0.09}_{-0.17}$ and  $\beta = -0.22^{+0.16}_{-0.29}$. These results establish lower bounds on the tidal charge parameter, namely $\beta_{min} = -0.3$ and $\beta_{min} = -0.51$, with a confidence level of $90\%$. 
To further assess the validity of the braneworld hypothesis relative to the Kerr model in GR, we calculate the Bayes factor between these two hypotheses, which informs us about the model that receives stronger support from the available data. 
For the two cases above, we find the corresponding log Bayes factor to be $ln\, \mathcal{B}^{\rm Kerr}_{\rm BW} = 0.77$ and $0.58$ respectively. This observation clearly suggests that there is no substantial evidence either in favor of or against the braneworld BH hypothesis from GW150914 ringdown data. We show these results in Table \ref{gwtc_2_table}.
The mild preference of Kerr vs the BW model is due to the degeneracy between the remnant spin and tidal charge parameters.
A large fraction of the beyond-Kerr parameter space remains allowed, preventing the Bayes Factor to decisively favour one of the two models.
This is in agreement with past literature on the U(1) charged case, see Table~I of Ref.~\cite{Carullo:2021oxn}.\\

\textit{Extended GWTC-3  Analysis  ---}  
To further test the braneworld hypothesis, we extend our analysis beyond GW150914 and explore a broader set of gravitational-wave events from GWTC-3.
Given the inconclusive results (as indicated by non-informative Bayes factors) from our initial analysis of GW150914, it is important to test the braneworld hypothesis with other events.
The results are presented in Figs.~\ref{GWTC_2_results},\ref{q_af_2d_posterior}, and Table~\ref{gwtc_2_table} for 10 selected events from GWTC-3 showing a loud ringdown signal. 
In Fig.~\ref{GWTC_2_results}, we show the tidal charge posterior samples for $t_0 = 0\, M_f$ (left panel) and $t_0 = 10\, M_f$ (right panel) after the peak. 
To provide a more detailed depiction of the correlation between spin and tidal charge, Fig.~\ref{q_af_2d_posterior} displays the 2D posterior samples of these parameters for four events. 
The scattered points represent the posterior samples, with their positions on the plot corresponding to specific ($\chi, \beta$) pairs. 
The color bar accompanying the plot indicates the log-likelihood associated with each of these pairs.
As can be deduced from these figures and Table~\ref{gwtc_2_table}, extreme values of the tidal parameter can be confidently excluded, with Bayes Factors typically mildly favouring the GR hypothesis.

\begin{table}[!tb]
\vspace{0.2cm}
\renewcommand{\arraystretch}{1.5} 
\setlength{\tabcolsep}{20pt} 
\resizebox{0.9\columnwidth}{!}{
\Huge
  \begin{tabular}{@{}lcccc} 
    
    \hline
    \hline
    Event & $\beta (0\, M_f)$ & $\ln B^{\mathrm{Kerr}}_{\mathrm{BW}}$ & $\beta (10\, M_f)$ & $\ln B^{\mathrm{Kerr}}_{\mathrm{BW}}$ \\
    \hline\\
    GW150914 & $-0.38$ & 0.77 & -0.62 & 0.58\\
    GW170104 & $-0.51$ & 0.66 & -0.61 & 0.45 \\
    GW170729 & $-0.54$ & 0.81 & -0.68 & 0.56 \\
    GW170814 & $-0.59$ & 0.46 & -0.67 & 0.29 \\
    GW170823 & $-0.52$ & 0.62 & -0.72 & 0.67 \\
    GW190521 & $-0.44$ & $0.62$ & -0.64 & 0.65 \\
    GW190521\_074359 & $-0.51$ & 0.85 & -0.60 & 0.53 \\
    GW190602\_175927 & $-0.65$ & 0.23 & -0.71 & 0.29 \\
    GW190706\_222641 & $-0.53$ & -0.14 & -0.73 & -0.16 \\
    GW190727\_060333 & $-0.66$ & 0.21 & -0.76 & 0.27 \\
    \hline\hline
    
  \end{tabular}
  
}
\caption{Lower bounds on the tidal charge parameter at $90 \%$ credibility, for two different start times, and corresponding Bayes Factors.}
\label{gwtc_2_table}
\end{table}

\section{Conclusions}\label{Section 5}

We considered the dynamical relaxation process of black holes in braneworld scenarios, computing their quasinormal spectrum, and constructing a waveform emission template capable of describing the final stages of the gravitational wave signal emitted by merging black holes.
In this class of theories, black holes possess a tidal charge, allowing the black hole angular momentum to cross the Kerr extremal bound without developing naked singularities.
This charge parameter bears direct imprints on the quasinormal spectrum of the black hole, giving rise to interesting phenomenology explorable with gravitational wave signals.
We applied this template to compact binary triggers detected by the LIGO-Virgo interferometers, employing a custom implementation of a Bayesian time-domain method.
Our analysis improves over the previous estimates presented in Ref.~\cite{Mishra:2021waw}, which ignored the correlations between the intrinsic black hole parameters and the tidal charge.
The state-of-the-art \texttt{dynesty} stochastic sampler was used to compute posterior probability distributions and the evidence between the GR and braneworld hypotheses.
We find lower bounds on adimensional tidal charges of $\sim -0.38$, excluding a significant portion of the parameter space.
Bayes Factors are only mildy informative, with a small preference for the GR hypothesis.

\section*{Acknowledgements}

We thank Abhirup Ghosh for his collaboration in the early stages of this work, the members of the `Testing General Relativity' group within the LIGO-Virgo-Kagra collaboration, and the astrophysical relativity group at ICTS for various stimulating discussions. We also appreciate the valuable comments on the manuscript provided by Nicola Franchini and Harrison Siegel.
Research of AKM is supported by SERB, Government of India through the National Post Doctoral Fellowship grant (Reg. No. PDF/2021/003081).
GC acknowledges funding from the European Union’s Horizon 2020 research and innovation program under the Marie Sklodowska-Curie grant agreement No. 847523 ‘INTERACTIONS’, from the Villum Investigator program supported by VILLUM FONDEN (grant no. 37766) and the DNRF Chair, by the Danish Research Foundation.
SC acknowledges the Albert Einstein Institute for warm hospitality, where a part of this work has been done, and also to Max-Planck Society for providing the Max-Planck-India mobility grant.  
Computations were performed with the aid of the Sonic HPC cluster at ICTS-TIFR.
Contents of this manuscript have been derived using the publicly available \texttt{python} software packages: \texttt{dynesty, lalsuite, matplotlib, numpy, pyRing, scipy}~\cite{dynesty, lalsuite, pyRing}.

\bibliography{References}

\bibliographystyle{./utphys1}
\end{document}